# Fast Inventory for 3GPP Ambient IoT Considering Device Unavailability due to Energy Harvesting

Zhikun Wu, Kazuk Takeda, Piyush Gupta, Ruiming Zheng, Luanxia Yang, Chengjin Zhang, Zhifei Fan, Hao Xu, Kiran Mukkavilli, and Tingfang Ji

*Abstract*— **With the growing demand for massive internet of things (IoT), new IoT technology, namely ambient IoT (A-IoT), has been studied in the 3rd Generation Partnership Project (3GPP). A-IoT devices are batteryless and consume ultra-low power, relying on energy harvesting and energy storage to capture a small amount of energy for communication. A promising use-case of A-IoT is inventory, where a reader communicates with hundreds of A-IoT devices to identify them. However, energy harvesting required before communication can significantly delay or even fail inventory completion. In this work, solutions including duty cycled monitoring (DCM), device grouping and low-power receiving chain are proposed. Evaluation results show that the time required for a reader to complete an inventory procedure for hundreds of A-IoT devices can be reduced by 50% to 83% with the proposed methods.**

## I. INTRODUCTION

THERE has been a growing demand for massive connectivity for internet of things (IoT). The 3rd Generation Partnership Project (3GPP) has conducted a feasibility study on design targets for new IoT technology that relies on ultra-low complexity devices with ultra-low power consumption for very low-end IoT applications in Release 18 [1]. The IoT devices are intended to be batteryless and equipped with limited energy storage (e.g., capacitors) that can be energy-harvested from ambient and external sources. These devices are referred to as ambient IoT (A-IoT) devices in 3GPP. Based on the outcome of the study in Release 18, 3GPP has further conducted a study of solutions for A-IoT in Release 19 [2]. The study outcome is summarized in the 3GPP technical report (TR) [3]. The outline of the 3GPP A-IoT framework is introduced below.

**(1) Device architectures**: 3GPP has studied two device types – device 1 with $1\mu W$ peak power consumption and device 2 with a few hundred $\mu W$ peak power consumption. Device 1 is a pure passive device without signal amplifiers or carrier wave (CW) generators, relying on backscattering an external CW, like UHF RFID ISO18000-6C [4]. Device 2 has active components such as amplifiers and an internal CW generator to achieve better performance than device 1 at the cost of increased complexity.

**(2) A-IoT reader**: Both base station (BS) and user equipment

(UE) have been studied in 3GPP as readers for A-IoT. For external CW used in backscatter communication of A-IoT devices, 3GPP considered scenarios where the CW is provided by the reader or by another node.

**(3) Use-cases/services**: Although various use-cases such as inventory, sensors, positioning, and command in indoor and outdoor environments have been identified, the 3GPP study of A-IoT solutions in Release 19 [2-3] focused on indoor inventory and indoor command, where inventory is an operation triggered by a reader to discover and acquire the identifiers of A-IoT devices, and command is an operation triggered by a reader to send the operation instruction to the A-IoT device (e.g., read, write, etc.)

**(4) Spectrum**: 3GPP identified operator-licensed FDD spectrum for A-IoT communications. Use of either the downlink carrier or uplink carrier of an FDD spectrum is considered depending on the reader type (BS or UE), communication direction (reader-to-device, device-to-reader), the frequency for external CW for backscatter communication devices, etc.

**(5) R2D (reader-to-device) communication**: As a unified design for all the device architectures, 3GPP identified OOK as the modulation scheme for R2D communication. A-IoT devices can implement either an RF envelope detector or a mixer-based receiver. A reader may re-use an NR OFDM transmitter to generate OOK waveform for R2D transmission. The use of line coding such as Manchester coding or pulse-interval encoding (PIE) is considered to help the device with R2D detection.

**(6) D2R (device-to-reader) communication**: Backscatter communication is enabled by switching impedance to modulate externally provided CW. The resulting D2R waveform in baseband is a sequence of OOK or BPSK symbols. Active transmission with internal CW generation, in theory, can modulate phase, amplitude, and/or frequency of the CW. Several modulation schemes have been studied and listed in the 3GPP TR [3] for backscatter/active communications. Unlike UHF RFID ISO18000-6C [4], 3GPP A-IoT considers the adoption of convolutional coding for D2R. Further, frequency-domain multiple access (FDMA) for D2R among A-IoT devices for an inventory or a command for a reader is considered feasible. For backscatter communication, FDMA can be enabled by small frequency shifts realized by using

Zhikun Wu, Ruiming Zheng, and Luanxia Yang are with QUALCOMM Wireless Communication Technologies (China) Limited, Beijing, 100013, China (e-mail: {zhikwu, rzheng, luanyang}@qti.qualcomm.com).

Kazuk Takeda is with Qualcomm Japan GK, Tokyo, 107-0062, Japan (e-mail: ktakeda@qti.qualcomm.com).

Piyush Gupta, Chengjin Zhang, Zhifei Fan, Kiran Mukkavilli, and Tingfang Ji are with Qualcomm Technologies, Inc, San Diego, CA 92121-1714, USA (e-mail: {pigupta, chengjin, zhifeif, kmukkavi, tji}@qti.qualcomm.com).

Hao Xu is with QUALCOMM International Inc, Beijing, 100013, China (e-mail: hxu@qti.qualcomm.com).

different frequencies of backscattering.

**(7) Performances**: The 3GPP TR [3] captures achievable link budget between a reader and an A-IoT device based on various radio parameters and identified the maximum distance targets of device 1 and device 2 as 10-15m and 15-50m, respectively. It was identified that a BS reader can achieve better link budget compared to a UE reader due to its higher transmit power for R2D and better CW cancellation capability for backscattered D2R. Among devices, device 2 with active components can achieve a better link budget than device 1 with purely passive components.

**(8) Device unavailability due to energy harvesting**: In the 3GPP study, it was identified that A-IoT devices may be unavailable for communication due to a lack of energy in storage, which results in a loss of spectral efficiency and/or increased latency. Unlike smartphones, A-IoT devices require frequent energy harvesting. If an A-IoT device does not have enough energy for A-IoT communication with a reader when the reader is about to communicate, the A-IoT device will be unavailable until the energy storage is charged. 3GPP studied various solutions for this and captured them in the TR [3].

This paper focuses on indoor inventory for A-IoT. It highlights that device unavailability due to energy harvesting significantly impacts the time required for the completion of inventory for A-IoT devices. The paper proposes a solution to reduce the completion time by addressing device unavailability caused by energy harvesting. Computer simulations demonstrate that the proposal reduces the inventory completion time by 50% to 83%.

## II. SYSTEM MODEL

For an A-IoT inventory, a reader triggers A-IoT random access procedure(s) to identify A-IoT device(s) in the reader's coverage. In this section, the A-IoT random access procedure and the issues of device unavailability caused by energy harvesting are presented.

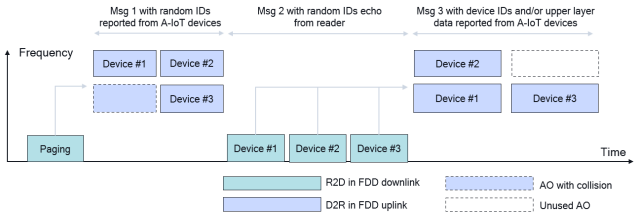

**Fig. 1.** A-IoT CBRA procedure and time-frequency arrangement

### A. A-IoT random access procedure

A-IoT paging is an R2D message triggering a random access procedure for one/multiple/all A-IoT devices. A-IoT random access can further be classified into contention-free random access (CFRA) and contention-based random access (CBRA). For A-IoT inventory for an unknown number of A-IoT devices, CBRA is usually used.

Steps of CBRA for A-IoT are modeled as in Fig. 1. In this model, an A-IoT device that received A-IoT paging generates a short random ID (e.g., 16 bits) and responds with an A-IoT Msg1 with the generated random ID. The A-IoT Msg1 is based on slotted-ALOHA, i.e., time/frequency resources for A-IoT Msg1, defined as access occasions (AOs), do not overlap each other, and each A-IoT device randomly selects one of the AOs for its A-IoT Msg1 transmission. If multiple A-IoT devices select the same AO, the CBRA will fail due to the collision. The reader that has triggered CBRA monitors AOs for potential A-IoT Msg1 reception(s), and if one or multiple A-IoT Msg1s are successfully decoded, the reader transmits one or multiple A-IoT Msg2s, each of which carries the echo of the random ID for the corresponding A-IoT Msg1. If an A-IoT device receives an A-IoT Msg2 containing the random ID that the device has transmitted in the A-IoT Msg1, the A-IoT device further reports the device ID and/or upper layer data in an A-IoT Msg3 which is scheduled by the A-IoT Msg2.

There can be a large number of A-IoT devices around a reader for an inventory. In order to complete an inventory for all the A-IoT devices, the reader may need to transmit A-IoT paging multiple times, in a periodic or an aperiodic manner, to trigger multiple CBRAs, until all the A-IoT devices are considered identified. It is assumed that an A-IoT device that has successfully transmitted its device ID via A-IoT Msg3 in a CBRA does not join future CBRAs triggered by the reader during the inventory. The necessary time for completing the inventory for a reader is increased due to CBRA failures, which can be caused by A-IoT Msg1 collisions, erroneous reception of A-IoT CBRA messages due to noise/interference, or lack of energy in the energy storage.

### B. Time and frequency resource arrangement for CBRA

The time and frequency resource arrangement for CBRA messages is illustrated in Fig. 1. For A-IoT Msg1 and Msg3, A-IoT may support frequency domain and time domain multiple access. Fig. 1 illustrates an example where an A-IoT paging indicates 4 AOs for A-IoT Msg1, with 2 AOs in the time domain and 2 AOs in the frequency domain. Each A-IoT device randomly selects one AO from the 4 AOs for its A-IoT Msg1 transmission. Then, the reader detects 3 A-IoT Msg1s in 3 of the 4 AOs and transmits 3 A-IoT Msg2s to respond and schedule A-IoT Msg3s. There is an A-IoT Msg1 collision in one AO. For this, the reader does not transmit the corresponding A-IoT Msg2. A-IoT devices, whose A-IoT Msg1s collide on the AO do not receive corresponding A-IoT Msg2s, and hence need to monitor another paging for another opportunity of CBRA.

### C. A-IoT device energy storage and energy harvesting models

Maximal energy that can be stored in the energy storage is denoted by $E_{es}^{max}$, and the instantaneous energy that is stored in the energy storage at a time is denoted as $e_{es}$, where $0 \leq e_{es} \leq E_{es}^{max}$. Additionally, it is assumed in this work that the reader transmits RF constantly for A-IoT devices to harvest energy, and an A-IoT device can harvest energy from that to the energy storage at a constant rate, which is denoted by $P_{eh}$, when the A-IoT device is in the "off state" or "sleep state", where the states will be presented in Section II-D and III-A, respectively. Let $p_{in}$ denote incident RF signal power at the A-IoT device, and $\xi(p_{in})$ denote power conversion efficiency of the A-IoT device RF energy harvesting. Then, $P_{eh} = p_{in}\xi(p_{in})$.

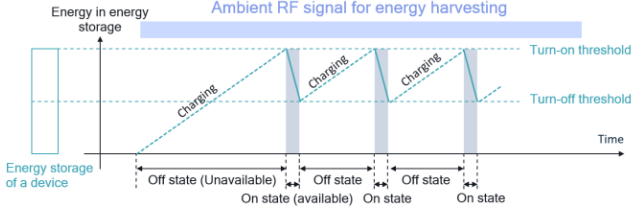

**Fig. 2.** $e_{es}$ with A-IoT device on/off states

### D. Conventional monitoring mechanism of A-IoT devices

In this paper, two states, namely "on state" and "off state" for A-IoT devices, are introduced to model the device availability for A-IoT communications and unavailability due to energy harvesting. An A-IoT device is in the on state when it is available for A-IoT reception and transmission. The power consumption of A-IoT device reception and transmission are denoted by $P_{rx}$ and $P_{tx}$, respectively. In the on state, $e_{es}$ depletes with the power consumption of A-IoT device reception or transmission. An A-IoT device is in the off state when the integrated circuit (IC) is turned off due to lack of energy, and hence, is unavailable for A-IoT reception and transmission. The power consumption of the off state, $P_{off}$, is negligible. An A-IoT device in the off state can harvest energy.

Fig. 2 shows how the state transition and the amount of energy $e_{es}$ for an A-IoT device in time look like when there is a constant RF signals for energy harvesting. As depicted in Fig. 2, when $e_{es}$ reaches the turn-on threshold $E_{es}^{up}$ ($<= E_{es}^{max}$), the A-IoT device transitions from the off state to the on state. Conversely, when $e_{es}$ falls below the turn-off threshold $E_{es}^{low}$ ($< E_{es}^{up}$), the A-IoT device enters the off state. The transition between the on state and the off state is entirely determined by the available energy in the energy storage. This mechanism is referred to as an energy-based monitoring (EM) mechanism in this work. Assuming an A-IoT device monitors A-IoT message reception from a reader with the power consumption of $P_{rx}$, then the maximum duration that an A-IoT device can monitor a channel in the on state ,denoted by $T_{on}^{EM}$, equals to ($E_{es}^{up} - E_{es}^{low}$)/$P_{rx}$.

### E. Problem

Since some A-IoT devices may be fully discharged and in the off state, the reader must send RF signals long enough for energy harvesting before starting inventory with A-IoT paging message transmissions. Therefore, an inventory process for a reader must consist of two stages, namely a charging stage and an inventory stage. In the charging stage, the reader sends RF signals for energy harvesting and does not communicate with A-IoT devices. In the inventory stage, the reader transmits one or more paging messages to trigger one or more CBRAs, where the number of paging messages in the inventory stage depends at least on the number of A-IoT devices for the inventory. The duration of the charging stage may depend on the expected time duration for a device with the smallest possible $P_{eh}$ among the devices in the coverage. This basically cannot be improved by A-IoT system design. In the inventory stage, the reader engages in communication with A-IoT devices for inventory. The duration of the inventory stage, denoted by $T_{inv}$, depends on the efficiency of the inventory procedure. Note that even during the inventory stage, the reader should continue providing RF signals for energy harvesting so that devices can charge whenever necessary or possible.

In the EM mechanism, the following three main problems can arise:

### (P1) need for a long time for energy harvesting even after the inventory stage starts

Different A-IoT devices have different amounts of energy in their storages at a time. Furthermore, some A-IoT devices are in the on state consuming power for monitoring paging, and some other A-IoT devices are in the off state harvesting energy with various $P_{eh}$. In the worst case, an A-IoT device may be in the off state with $e_{es}$ being close to $E_{es}^{low}$ when a reader just starts the inventory stage. Such an A-IoT device needs energy harvesting until the energy in the storage reaches $E_{es}^{up}$ even after the reader starts the inventory stage. During the energy harvesting, the A-IoT device misses paging. This would take a long time and cause a large delay for inventory completion for the reader. For instance, for the A-IoT device, if the energy to harvest from $E_{es}^{low}$ to $E_{es}^{up}$ is 250nJ , and if $p_{in} = -36dBm$ and $\xi(p_{in}) = 5\%$, the A-IoT device needs nearly 20 seconds to be in the on state. In other words, in the first 20 seconds of the inventory stage of the reader, the A-IoT device is not available for communication .

### (P2) lack of energy to conduct the CBRA triggered by a paging

When being paged, an A-IoT device may be in the on state but its energy level $e_{es}$ is approaching $E_{es}^{low}$. Such an A-IoT device would not be able to complete the process of the CBRA due to insufficient energy. If $e_{es}$ reaches $E_{es}^{low}$, the device needs to harvest energy from $E_{es}^{low}$ to $E_{es}^{up}$, in order to recover for paging monitoring and CBRA.

### (P3) need for several re-accesses for CBRA due to congestion

Even if the energy in the storage is high, A-IoT Msg1 collisions or erroneous reception of A-IoT messages in CBRA can cause CBRA failures. In such cases, an A-IoT device may need to re-access multiple times. However, the available energy in the energy storage may not be sufficient to sustain long-time device activities for such multiple CBRAs. Assume the available amount of energy in the storage for the device activities in the on state, ($E_{es}^{up} - E_{es}^{low}$), is 500nJ. With $P_{rx} = 50uW$, the A-IoT device is available for at most 50ms. Assuming at most 5 A-IoT paging messages (no matter periodic or aperiodic) are transmitted in 50ms, at most 5 CBRAs are available for the A-IoT device during the 50ms. If the device fails to access/re-access over the 5 CBRAs due to e.g., A-IoT Msg1 collisions, the A-IoT device will have to switch to the off state and needs to harvest energy from $E_{es}^{low}$ to $E_{es}^{up}$, in order to recover for paging monitoring and CBRA.

### III. Proposed DCM and congestion control mechanism

To address those problems, a new mechanism, namely duty cycle monitoring (DCM), is proposed in this work. Additionally, congestion control related solutions, namely access probability adjustment and device grouping are also proposed for scenarios where a reader performs inventory for many A-IoT devices in this section.

#### A. DCM

DCM [5] is a mechanism that controls the duration of the on state, so that the time duration of the off state for energy harvesting can be shortened/controlled. Before an A-IoT device detects a paging trigger for a CBRA, A-IoT devices monitor the channel with an additional timer applied in the on state. The timer, namely the "on timer", is used for restricting the on duration $T_{on}^{timer}$ to be shorter than $T_{on}^{EM}$ if no paging is received from the reader. The duration $T_{on}^{timer}$ can be determined by the device implementation. However, if A-IoT paging is transmitted with a fixed periodicity $T_{pg}$, it is beneficial to design the duration of the on timer such that $T_{on}^{timer} \geq T_{pg}$ so that at least one A-IoT paging can be received in the on timer duration when it comes to the inventory stage. Once the energy in the storage reaches the turn-on threshold $E_{es}^{up}$, the A-IoT device monitors the channel for paging for a duration of $T_{on}^{timer}$. When the on timer expires without receiving any paging, A-IoT devices transition to the off state and harvest energy until the energy in the storage reaches $E_{es}^{up}$. Other features remain unchanged compared to the EM mechanism. With $T_{on}^{timer}$ being much shorter than $T_{on}^{EM}$, A-IoT devices maintain higher level of $e_{es}$. As shown in Fig. 3, with limited energy consumption during $T_{on}^{timer}$, (1) even if an A-IoT device is in the off state when the reader starts the inventory stage, the time required for harvesting energy for charging until $E_{es}^{up}$ is much shorter than with the EM mechanism, which resolves **P1**; (2) when an A-IoT device joins the inventory based on a paging, the energy in the energy storage is guaranteed to be high, which resolves **P2**.

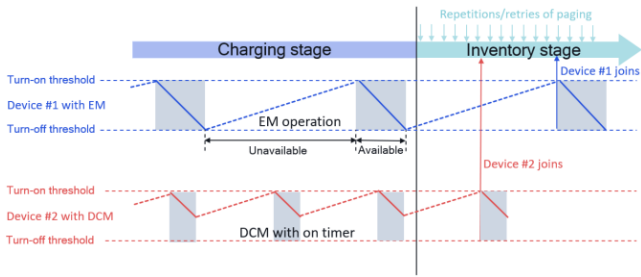

**Fig. 3.** Comparison of EM mechanism and DCM mechanism.

Once the A-IoT device with DCM detects a paging, it identifies that there is a reader conducting an inventory stage. If an A-IoT device receives a paging, the device proceeds with further message exchanges with the reader for the current CBRA, and if re-access is necessary due to CBRA failure, the device further proceeds with subsequent CBRAs. In the DCM mechanism, an A-IoT device after acquiring the inventory stage for a reader is assumed to control the duration and timing of the on state by using a new state, the sleep state. In the sleep state, an A-IoT device can turn on the IC to maintain a sleep timer that can count the time until the next occasion when the A-IoT device has to be in the on state for the message exchange for the current or subsequent CBRAs. Similar to the off state, an A-IoT device in the sleep state can harvest energy but is not able to transmit or receive. Let $T_{sl}^{DCM}$ and $T_{on}^{DCM}$ denote the sleep duration and on duration for monitoring paging in the DCM mechanism, respectively. By coordinating the sleep timer and the on timer, the DCM mechanism ensures that $T_{sl}^{DCM} + T_{on}^{DCM} = T_{pg}$. In other words, whenever an A-IoT device transitions to the on state, there is a paging transmitted from the reader. For this, synchronization is crucial. Such synchronization is possible if a synchronization signal is provided by the reader, e.g., as a reference signal in the paging. For simplicity, it is assumed in this paper that the A-IoT device is synchronized with the reader after receiving the very first A-IoT paging in the inventory stage. The power consumption during the sleep state is denoted by $P_{sl}$, satisfying $P_{sl} < P_{rx}$. Thanks to the power reduction in the sleep state compared to the on state, the available time of A-IoT device can be significantly increased. Thus, **P3** can be solved.

#### B. Access probability adjustment for congestion control

In the 3GPP study, a reader is supposed to perform inventory for hundreds of A-IoT devices. If many of these A-IoT devices access CBRA based on the same paging, serious congestion on the A-IoT Msg1 can occur. To solve this problem, the reader should be able to control the congestion. More specifically, an A-IoT paging can indicate an access probability for the paging, which is the probability that an A-IoT device receiving the A-IoT paging will transmit A-IoT Msg1. The reader is able to decide and indicate the access probability in a paging according to the observation of occupancy or congestion level of A-IoT Msg1 AOs in the CBRA(s) triggered by previous paging(s). By this method, the number of A-IoT devices responding to each paging can be well controlled, and the collision probability of A-IoT Msg1 can be reduced.

#### C. Device grouping for congestion control

As presented in Section III-B, the access probability adjustment for congestion control can reduce the probability of A-IoT Msg1 collisions among A-IoT devices for a given paging. However, this means that many A-IoT devices may monitor the same paging but not respond to it with high probability. Such an A-IoT device only consumes energy due to paging monitoring without CBRA, which may quickly drain the A-IoT device energy storage. To solve this problem, the sleep timer can be designed such that an A-IoT device wakes up with periodicity $N_g T_{pg}$, where $N_g$ is a non-zero integer value. For instance, when $N_g = 2$, an A-IoT device that receives odd paging from the reader continues monitoring odd paging from the reader using DCM; similarly for A-IoT devices that receive even paging, as illustrated in Fig. 4.

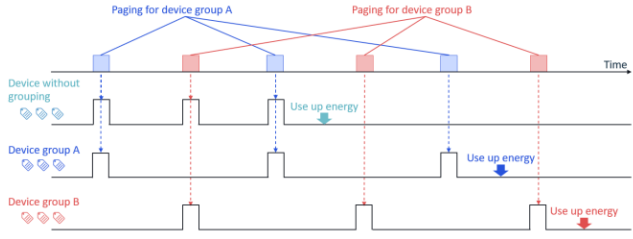

**Fig. 4.** Illustration of device grouping

### D. Low-power wake up receiver

In both the EM mechanism and the DCM mechanism, an A-IoT device monitors paging in the on state until it detects a paging from a reader. For reliable detection, the R2D transmission for paging is associated with a preamble that uses a particular OOK sequence known by the device in advance. The known sequence detection allows a device to use a low-power wake up receiver for the R2D detection. The use of a low-power wake up receiver is particularly beneficial for device 2, whose power consumption for A-IoT communication is much higher than that of device 1. For example, device 2 may implement an RF envelope detector based low-power wake up receiver [6] and use it for R2D detection without activating the main IC for A-IoT communication until it detects a paging. This contributes to device power saving for R2D monitoring, at the cost of R2D detection performance.

## IV. EVALUATION

In this section, simulation results are provided to validate the analysis and evaluate the performance of the proposed DCM mechanism with congestion control methods. Both device 1 and device 2 with different energy storage sizes and power consumptions [7] are evaluated. The evaluation parameters are summarized in Table 1.

### TABLE I
### EVALUATION PARAMETERS

| Parameters | Device 1 value | Device 2 value |
|---|---|---|
| Number of tags, $N$ | 600 | |
| Energy storage size, $E_{es}^{max}$ | 500nJ | 5000nJ |
| The turn-on threshold, $E_{es}^{up}$ | $E_{es}^{max}$ | |
| The turn-off threshold, $E_{es}^{low}$ | $0.5E_{es}^{max}$ | |
| Power consumption for reception, $P_{rx}$ | 1uW | 50uW |
| Power consumption for transmission, $P_{tx}$ | 1uW | 200uW |
| Power consumption for sleep state, $P_{sl}$ | 0.1uW | |
| Power consumption for low power wake up receiver | 1uW | |
| Duration of a time unit, slot | 0.5ms | |
| Paging duration | 2 slots | |
| Periodicity of paging (if periodic) $T_{pg}$ | 24 slots | 28 slots |
| A-IoT Msg1 duration | 1 slot | |
| A-IoT Msg2 duration | 1 slot | |
| A-IoT Msg3 duration | 6 slots | |
| DCM on duration $T_{on}^{DCM}$ after the device acquires the inventory stage of a reader | 4 slots | 2 slots |
| DCM on duration $T_{on}^{timer}$ before the device acquires the inventory stage of a reader | 36 slots | 52 slots |
| Number of AOs for A-IoT Msg1 in the time domain | 4 | |
| Number of AOs for A-IoT Msg1 in the frequency domain | 2 | 4 |

In the evaluation, the value of $p_{in}$ for each A-IoT device is based on the A-IoT device geographical distribution around the reader. More specifically, as described in the D1T1 layout of the TR [3], A-IoT devices are assumed to be distributed in a 120m x 60m indoor factory, and a base station, from the 18 base stations deployed within the factory, works for inventory at a time with its transmission power of 33 dBm. In the layout, the cumulative distribution function (CDF) of $p_{in}$ is as shown in Fig. 5 (a). Here, A-IoT devices with $p_{in}$ smaller than the receiver chain sensitivity of -36dBm are not accounted for. As for power conversion efficiency $\xi(p_{in})$, because $\xi(p_{in})$ is not a monotonically increasing function of $p_{in}$ [8], the following function is used in the evaluation, where $p_{in}$ unit is in dBm.

$$\xi(p_{in}) = \begin{cases} (p_{in} + 41)/100 & p_{in} \geq -10dBm \\ (-2p_{in} + 11)/100 & p_{in} < -10dBm \end{cases}$$

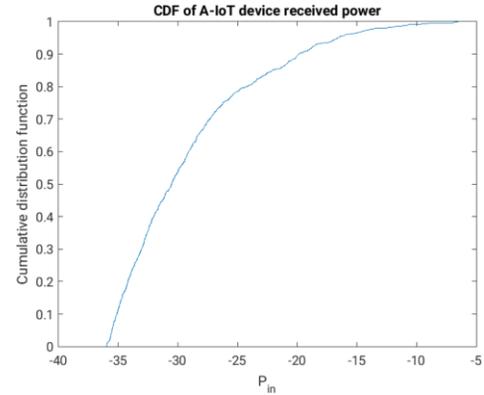

(a)

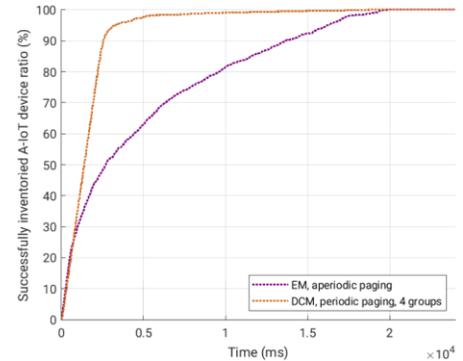

(b)

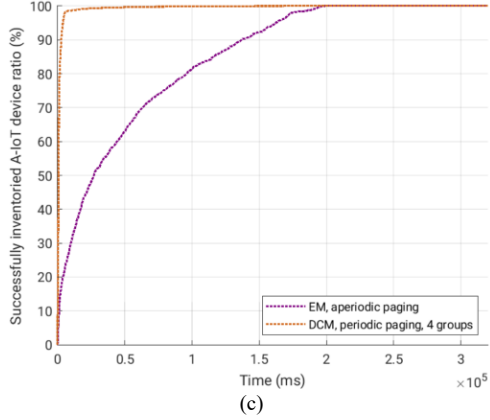

(c)

**Fig. 5.** (a) CDF of $p_{in}$ (dBm) in the layout; (b) ratio of A-IoT devices that have successfully been inventoried in time (device 1); (c) ratio of A-IoT devices that have successfully been inventoried in time (device 2)

Fig. 5 (b) illustrates the ratio of A-IoT devices that have not yet successfully completed CBRA at a given time. All the A-IoT devices are assumed to be device 1. For the EM mechanism, nearly 20 seconds are needed to successfully complete CBRAs for 99% of A-IoT devices. This is because of **P1** and **P2** in Section II-E. For the A-IoT devices whose $e_{es}$ is close to $E_{es}^{low}$ and $p_{in} \sim -36dBm$, nearly 20 seconds are necessary for energy harvesting for $(E_{es}^{up} - E_{es}^{low})$. However, with DCM, $e_{es}$ at the beginning of the inventory stage is not smaller than (428nJ) due to the limitation in on duration, which means they only need at most 5.76s to harvest energy and switch to the on state from the sleep state. The A-IoT devices with DCM can be quickly fully charged and join the inventory. Therefore, DCM can enable the reader to finish 99% of A-IoT devices inventory in 10 seconds. Fig. 5 (c) depicts the results of A-IoT device type 2. The tendency is similar to A-IoT device type 1. Compared to the EM mechanism, DCM can reduce inventory time overhead by 83%.

## V. CONCLUSIONS

In this paper, indoor inventory for A-IoT devices with energy harvesting is studied. As device unavailability due to energy harvesting significantly increases inventory completion time, solutions, including DCM and device grouping, have been proposed. Those solutions control energy in the storage and device availability/unavailability, thereby avoids the failure of an inventory for an A-IoT device due to energy depletion. Evaluation results have shown the effectiveness of the proposed schemes. The proposed schemes have been captured in the 3GPP TR [3] as a solution for device availability/unavailability due to energy harvesting.

## REFERENCES

[1] 3GPP, Technical Report 38.840, V18.0.0, Study on Ambient IoT (Internet of Things) in RAN, 2023.

[2] 3GPP, RP-234058, New SID: Study on solutions for Ambient IoT (Internet of Things) in NR, Huawei, 2023.

[3] 3GPP, Technical Report 38.769, V2.0.0, Study on solutions for ambient IoT (Internet of Things), 2024.

[4] EPC Radio-Frequency Identity Protocols, Class-1 Generation-2 UHF RFID Protocol for Communications at 860 MHz–960 MHz, Version 1.2.0, EPC Global, 2008.

[5] 3GPP, R1-2410480, Frame structure and timing aspects, Qualcomm Incorporated, 2024.

[6] 3GPP, Technical Report 38.869, Study on low-power wake up signal and receiver for NR, V18.0.0, 2023.

[7] 3GPP, R1-2410478, Ambient IoT Device Architecture, Qualcomm Incorporated, 2024.

[8] B. Clerckx, J. Kim, K. W. Choi and D. I. Kim, "Foundations of wireless information and power transfer: Theory, prototypes, and experiments," Proc. IEEE, vol. 110, no. 1, pp. 8–30, 2022.